\documentclass[amsmath,amssymb,aps,prl,twocolumn,superscriptaddress]{revtex4-2}

\usepackage{graphicx}
\usepackage{hyperref}
\usepackage{MnSymbol}
\usepackage[normalem]{ulem}

\DeclareMathOperator{\Prob}{Prob}
\DeclareMathOperator{\Tr}{Tr}
\DeclareMathOperator{\diag}{diag}
\DeclareMathOperator{\const}{const}

\begin{document}

\title{Measurement-induced phase transition for free fermions above one dimension}

\author{Igor Poboiko}
\author{Igor V. Gornyi}
\author{Alexander D. Mirlin}
\affiliation{\mbox{Institute for Quantum Materials and Technologies, Karlsruhe Institute of Technology, 76131 Karlsruhe, Germany}}
\affiliation{\mbox{Institut f\"ur Theorie der Kondensierten Materie, Karlsruhe Institute of Technology, 76131 Karlsruhe, Germany}}

\date{\today}

\begin{abstract}
A theory of the measurement-induced entanglement phase transition for free-fermion models in $d>1$ dimensions is developed. The critical point separates a gapless phase with $\ell^{d-1} \ln \ell$ scaling of the second cumulant of the particle number and of the entanglement entropy and an area-law phase with $\ell^{d-1}$ scaling, where $\ell$ is a size of the subsystem. The problem is mapped onto an SU($R$) replica non-linear sigma model in $d+1$ dimensions, with $R\to 1$. Using renormalization-group analysis, we calculate 
 critical indices in one-loop approximation justified for  $d = 1+ \epsilon$ with $\epsilon \ll 1$. Further, we carry out a numerical study of the transition for a $d=2$ model on a square lattice, determine numerically the critical point, and estimate the critical index  of the correlation length, $\nu \approx 1.4$. 

\end{abstract}

\maketitle


\textbf{\textit{Introduction}} ---
Quantum dynamics of many-body systems subjected to quantum measurements and, in particular, measurement-induced entanglement phase transitions attract a great deal of research attention. This area of research connects the condensed matter physics with quantum information processing. Measurement-induced transitions, which result from a competition between unitary dynamics that enhances the entanglement and measurements that reduce it, were 
 initially investigated in quantum-circuit framework \cite{Li2018a, Skinner2019a, Chan2019a, Szyniszewski2019a, Li2019a,  Bao2020a, Choi2020a, Gullans2020a, Gullans2020b, Jian2020a, Zabalo2020a, Iaconis2020a, Turkeshi2020a, Zhang2020c, Szyniszewski2020, Nahum2021a, Ippoliti2021a,    Ippoliti2021b,Lavasani2021a,Lavasani2021b,Sang2021a,Fisher2022,Block2022a,Sharma2022, 
 Agrawal2022, Barratt2022, Jian2023,Kelly2023}.
 Later works showed ubiquity of measurement-induced transitions; they were studied in a variety of many-body systems, including free fermions \cite{Cao2019a,Alberton2021a,Chen2020a,Tang2021a,Coppola2022,Ladewig2022,Carollo2022,Buchhold2022,Yang2022,Szyniszewski2022,Buchhold2021a,Buchhold2021a,VanRegemortel2021a,Youenn2023,Loio2023,Turkeshi2022b,Kells2023,Poboiko2023}, Majorana fermions \cite{Fava2023,Swann2023,Merritt2023}, Ising spin systems  \cite{Lang2020a,Rossini2020a,Biella2021a,Turkeshi2021,Tirrito2022,Yang2023,Weinstein2023,Murciano2023,Sierant2022a,Turkeshi2022a}, Bose-Hubbard models \cite{Tang2020a,Goto2020a,Fuji2020a,Doggen2022a,Doggen2023}, disordered systems with Anderson or many-body localization \cite{Szyniszewski2022,Lunt2020a,Yamamoto2023}, and models of Sachdev-Ye-Kitaev type \cite{Jian2021a,Altland2022}. While most of the studies were of computational character, important analytical progress has been achieved for some of the models.
 Furthermore, recent works on trapped-ion systems  \cite{Noel2022a, Agrawal2023} and superconducting quantum processors \cite{Koh2022, Hoke2023} reported experimental indications of
 measurement-induced phase transitions. 

 Recently, the present authors developed a field-theoretical approach to the measurement-induced physics in the model of one-dimensional free fermions \cite{Poboiko2023}. The problem was mapped onto a two-dimensional SU($R$) non-linear sigma-model (NLSM) in the replica limit $R\to 1$blue; see also Refs.~\cite{Jian2023, Fava2023} for related  NLSMs of different symmetries in Majorana fermion setups. The mapping to $\mathrm{SU}(R)$ NLSM allowed us to show that this system is always in the area-law phase but, for small measurement rate, exhibits a broad intermediate regime with logarithmic  increase of the entropy. These analytical results were supported by numerical simulations. 

 In the present work, we demonstrate that, in spatial dimensionality $d>1$, the system of free fermions subjected to measurements exhibits a phase transition between a gapless phase with $\ell^{d-1}\ln\ell$ scaling of the second cumulant of the particle number and of the entanglement entropy and an area-law phase with $\ell^{d-1}$ scaling. (Here $\ell$ is a size of the subsystem.) Specifically, on the analytical side, we map the problem onto a replica NLSM in $d+1$ dimensions and develop a renormalization group (RG) analysis. We show that for $d > 1$ and sufficiently rare measurements the scaling $\sim \ell^{d-1} \ln \ell$ holds. For $d = 1+\epsilon$ with $\epsilon \ll 1$ we develop a parametrically justified one-loop analysis of the transition and calculate critical indices. We identify the particle number covariance, closely related to the mutual information, as the scaling variable for the transition. Using this framework, we carry out a computational study of the transition in $d=2$ dimensions, determine numerically the critical point, and estimate the correlation-length critical exponent $\nu$. These arguments, together with the expectation of the quantum Zeno effect for sufficiently frequent measurements, imply the existence of the transition for arbitrary $d > 1$.


\textbf{\textit{Model and measurement protocol}} ---
We consider a $d$-dimensional tight-binding model of free fermions, defined on a hypercubic lattice with $L^d$ sites and periodic boundary conditions. The system Hamiltonian reads:
\begin{equation}
\hat{H}=-J\sum_{\left\langle \boldsymbol{x}\boldsymbol{x}^{\prime}\right\rangle }\left[\hat{\psi}^{\dagger}(\boldsymbol{x})\hat{\psi}(\boldsymbol{x}^{\prime})+h.c.\right],
\end{equation}
where summation is performed over lattice links and the hopping constant $J$ defines the energy scale. Each site $\boldsymbol{x}$ of the system is projectively monitored: at randomly chosen times $t_{i}^{(\boldsymbol{x})}$ sampled from a Poissonian distribution with a rate $\gamma$, independently for each site, a projective measurement of the site occupation number
$\hat{n}(\boldsymbol{x})=\hat{\psi}^{\dagger}(\boldsymbol{x})\hat{\psi}(\boldsymbol{x})$
is performed. The outcome of such measurements, which can be either zero or unity, $n=0,1$, is random, with probabilities determined by the many-body wave function $\Psi$ at time $t-0$: 
\begin{equation}
    \Prob(n|(\boldsymbol{x},t))=\left<\Psi(t-0)\right|\hat{\mathbb{P}}_n(\boldsymbol{x})\left|\Psi(t-0)\right>.
\end{equation}
This standard quantum-mechanical Born rule involves many-body projectors onto linear subspaces with a definite site occupation $n=0,1$: 
\begin{equation}
\hat{\mathbb{P}}_0(\boldsymbol{x})=1-\hat{n}(\boldsymbol{x}),\quad\hat{\mathbb{P}}_1(\boldsymbol{x})=\hat{n}(\boldsymbol{x}).
\end{equation}
After the measurement at time $t$ on site $\boldsymbol{x}$ with an outcome $n$ is performed, the wavefunction undergoes a von Neumann collapse and is projected onto the corresponding subspace with subsequent renormalization:
\begin{equation}
\left|\Psi(t+0)\right>=
\frac{\hat{\mathbb{P}}_{n}(\boldsymbol{x})\left|\Psi(t-0)\right> }{\left\Vert \hat{\mathbb{P}}_{n}(\boldsymbol{x})\left|\Psi(t-0)\right>\right\Vert }.
\end{equation}

The system is initially prepared in an arbitrary Gaussian pure state (i.e., described by a Slater determinant) $\left|\Psi(t_{0})\right>$. The evolution consists of unitary part governed by the Hamiltonian $\hat{H}$ and non-unitary part introduced by measurements according to the above protocol. We will study properties of the wavefunction $\left|\Psi(t=0)\right>$ in the limit $t_0\to-\infty$ (``steady state''). This pure state depends, by construction, on the random ``measurement trajectory'' ${\cal T}=\{t_m, \boldsymbol{x}_m, n_m\}$.

To investigate the measurement-induced entanglement transition, we utilize a relation \cite{KlichLevitov} between the entanglement entropy ${\cal S}_{E}$ for a given subsystem $A$ (characterized by the reduced density matrix $\hat\rho_A$) and an infinite series of particle-number cumulants, 
\begin{equation}
{\cal C}_{A}^{(N)}=\left\llangle \left(\sum_{\boldsymbol{x}\in A}\hat{n}(\boldsymbol{x})\right)^{N} \right\rrangle,
\end{equation}
in the same subsystem. 
This relation,
\begin{equation}
\label{eq:KlichLevitov}
{\cal S}_{E}\equiv-\Tr(\hat{\rho}_{A}\ln\hat{\rho}_{A})=\sum_{q=1}^{\infty}2\zeta(2q){\cal C}_{A}^{(2q)}
\end{equation}
with $\zeta$ the zeta-function,
holds for arbitrary pure Gaussian states, including~\cite{Turkeshi2022b, Szyniszewski2022, Poboiko2023} monitored systems. Moreover, it was demonstrated \cite{Poboiko2023} 
 that the first term (involving the second particle-number cumulant
  ${\cal C}_{A}^{(2)}$) is dominant in the series~\eqref{eq:KlichLevitov}, i.e., ${\cal S}_{E} \approx(\pi^2/3){\cal C}_{A}^{(2)}$. This approximation is parametrically controlled for small measurement rate $\gamma/J$ and holds numerically with an excellent precision even when $\gamma/J$ is not small. We have verified that this is the case also for $d=2$ across the transition \cite{SuppMat}.
  This property is also known to hold in disordered systems~\cite{Burmistrov2017}, even in the vicinity of the Anderson metal-insulator transition. Notably, the present problem bears a certain similarity to that of Anderson transitions~\cite{evers08}, as both can be mapped to corresponding NLSMs, albeit with different replica limits. For these reasons, our further analysis focuses on the second particle-number cumulant ${\cal C}_{A}^{(2)}$, with the results applying also to the entanglement entropy  ${\cal S}_{E}$. Note that, in interacting systems, the above relation between ${\cal S}_{E}$ and particle-number fluctuations does not hold, making it possible to have a transition in particle-number fluctuations (``charge sharpening'') distinct from the entanglement transition \cite{Agrawal2022,Barratt2022, Agrawal2023}.


\textbf{\textit{Non-linear sigma model}} --- In Ref.~\cite{Poboiko2023}, it was shown that the behavior of the particle-number cumulants (and of the entanglement entropy) in monitored one-dimensional free-fermion systems is described by a replicated NLSM. Specifically, this field theory operates with matrix fields ${\hat{U}(\boldsymbol{x}, t)\in \text{SU}(R)}$ in the replica space, 
governed by the Lagrangian density 
\begin{equation}
\label{eq:SU-R-action}
{\cal L}[\hat{U}]=\frac{g}{2}\Tr\left(\frac{1}{v_{0}}\partial_{t}^{\Xi}\hat{U}(\partial_{t}^{\Xi}\hat{U})^{\dagger}+v_{0}\nabla\hat{U}\nabla\hat{U}^{\dagger}\right),
\end{equation}
and yields the observable correlation functions in the 
replica limit $R \to 1$ (dictated by the Born rule). The derivation of the effective theory (\ref{eq:SU-R-action}) is straightforwardly extended to arbitrary spatial dimensionality $d$; as a result, we obtain the NLSM in $d+1$ dimensions \cite{SuppMat}. The theory is defined in the time domain $t \leq 0$, with the boundary condition $\hat{U}(\boldsymbol{x}, t = 0) = \hat{\mathbb{I}}$. As a consequence of the measurement-induced infinite-temperature heating of the system in the steady state, velocity in Eq.~\eqref{eq:SU-R-action} is expressed via the root-mean-square velocity averaged over the Brillouin zone $v = \sqrt{2d} J$ as $v_0 = v / \sqrt{d}$.
The bare value of the coupling constant $g$ is given by:
\begin{equation}
g_{0}=\rho(1-\rho)v_0/\gamma
\label{eq:8}
\end{equation}
with $\rho \in [0, 1]$ being an average filling factor of the state (which is conserved during the evolution). The time derivative in Eq.~(\ref{eq:SU-R-action}) contains a source matrix field $\hat{\Xi}(\boldsymbol{x},t)$, which is diagonal with respect to replica indices, $\hat{\Xi}=\diag(\{\Xi_r\}_{r=1}^R)$:
\begin{equation}
\label{eq:Density-source}
\partial_{t}^{\Xi}\hat{U}\equiv\partial_{t}\hat{U}+(i Z/2)\{\hat{U},\hat{\Xi}\}.
\end{equation}
Here $\{.,.\}$ is the anticommutator and we have also introduced a renormalization factor $Z$ with the bare value $Z_0=1$.

The generating functional $\mathcal{Z}[\hat{\Xi}]$ obtained from Eq.~\eqref{eq:SU-R-action} in the presence of the source is then used to calculate the density correlation function that determines the particle-number cumulants in Eq.~\eqref{eq:KlichLevitov}. The simplest,  two-point correlation function is given by
\begin{multline}
\!C(\boldsymbol{x})=\overline{\left\langle \left\{ \hat{n}(\boldsymbol{x}),\hat{n}(0)\right\} /2\right\rangle }-\overline{\left\langle \hat{n}(\boldsymbol{x})\right\rangle \left\langle \hat{n}(0)\right\rangle }\\
\!=\!-\!\!\lim_{\substack{t,t^{\prime}\to0\\
R\to1}}\frac{1}{R-1}\Bigg[\!\frac{g_{0}}{v_{0}}\delta(\boldsymbol{x})\delta(t-t^\prime) + \sum_{r=1}^{R}
\frac{\delta^{2}\ln \mathcal{Z}[\hat{\Xi}]}{\delta\Xi_{r}(\boldsymbol{x},t)\delta\Xi_{r}(0,t^\prime)}\!
\Bigg],
\label{eq:C-x-definition}
\end{multline}
where the overbar denotes averaging over quantum trajectories $\mathcal{T}$.
On the Gaussian level (i.e., expanding the fields in small fluctuations around the saddle point up to second order),
this correlation function and its Fourier transform read:
\begin{equation}
\label{eq:CqGaussian}
C(\boldsymbol{x})=-2 g_0 Z_0^{2} / \sigma_d |\boldsymbol{x}|^{d+1}, \quad C(\boldsymbol{q})=g_{0} Z_0^{2}|\boldsymbol{q}|, 
\end{equation}
where $\sigma_d = 2\pi^{(d+1)/2}/\Gamma(d/2+1/2)$ is the surface area of $d$-dimensional sphere (with $\Gamma$ the gamma function).  For a subsystem in the form of a  ($d$-dimensional) ball of radius $\ell$, this correlation function leads to the following behavior of the second cumulant in the Gaussian approximation:
\begin{equation}
\label{eq:Cumulant:Critical}
{\cal C}_{\ell}^{(2)}\simeq\frac{g_{0}}{\pi}\sigma_{d-1}\ell^{d-1}\ln\frac{\ell}{l_{0}},
\end{equation}
where $l_0 \simeq \sqrt{d/2} (J/\gamma)$ is the mean free path that plays a role of the ultraviolet cutoff for the NLSM. Importantly, a power-law (``critical'') character \eqref{eq:CqGaussian} of $C(\boldsymbol{x})$ leads to a logarithmic enhancement of ${\cal C}_{\ell}^{(2)}$ and of the entanglement entropy in comparison to area law.


\textbf{\textit{RG and scaling analysis}} --- The long-wavelength behavior of the NLSM~\eqref{eq:SU-R-action} can be analyzed by means of RG equations for two running dimensionless coupling constants, $G(\ell) = g(\ell)\ell^{d-1}$ and $Z(\ell)$, cf. Refs. \cite{HikamiRG,WegnerRG}. Since $d=1$ corresponds to the logarithmic dimension for the NLSM theory, we can develop an $\epsilon$-expansion with ${\epsilon=d-1}$. 
In the one-loop approximation, the RG flow is described by (cf. Ref.~\cite{Poboiko2023} where $\epsilon=0$; see also \cite{SuppMat}):
\begin{align}
dG / d\ln \ell&\equiv\beta(G)=\epsilon\, G-R/4\pi+O(1/G),
\label{betaG}\\
d\ln Z / d\ln \ell &\equiv\beta_{Z}(G)= O(1/G^{2}).
\label{betaZ}
\end{align}
Equation~\eqref{betaG} agrees  with earlier results for the $\beta$-function of the NLSM of corresponding symmetry~\cite{Konig2012,HikamiRG,WegnerRG,evers08}.

For spatial dimensions higher than unity, $\epsilon>0$, Eq.~\eqref{betaG} implies the existence of a critical point defined by the condition $\beta(G_c)=0$. Following an analogy with the Anderson metal-insulator transition, we will call the phase governed by an infrared-stable fixed point $G = 0$ ``insulating'', which  corresponds to the area-law phase in the context of measurement-induced phase transitions. The phase with $G\to\infty$ will be referred to as ``metallic,'' where the scaling of the cumulant with $\ell$ will be given by Eq.~\eqref{eq:Cumulant:Critical} with a renormalized prefactor. In the replica limit $R\to1$ and to leading order in ${\epsilon\ll 1}$, we find the critical point separating these two phases at
\begin{equation}
G_{c}\approx1/{4\pi\epsilon}+O(\epsilon^{0}),
\label{Gc}
\end{equation}
akin to the critical conductance in Anderson transitions \cite{Abrahams1979scaling,Kramer1993localization,Slevin1997Anderson,evers08,Shapiro1990}.

Expanding RG equations \eqref{betaG} and \eqref{betaZ} around the critical point, ${\beta(G)\approx\beta^\prime(G_c)(G-G_c)\equiv (G-G_c)/\nu}$ and ${\beta_Z(G)\approx\beta_Z(G_c)\equiv\zeta/2}$, we obtain the following one-loop results for the corresponding critical exponents:
\begin{align}
\nu	=1/\epsilon+O(\epsilon^{0}),\quad 
\zeta =0+O(\epsilon^{2}).
\label{eq:crit-exp-one-loop}
\end{align}
Integration of the RG equations in the vicinity of the critical point yields:
\begin{align}
G(\ell) - G_{c} \approx (G_{0}-G_{c})\cdot\left(\ell/l_{0}\right)^{1/\nu},\quad 
Z(\ell)\approx\left(\ell/l_{0}\right)^{\zeta/2}.
\label{eq:G_Z_critical-scaling}
\end{align}
The length scale at which ${|G-G_c|\sim G_c}$ defines the correlation length:
\begin{equation}
l_{\text{corr}}\sim l_{0}
(|G_{0}-G_{c}| / G_c)^{-\nu}.
\label{eq:l-corr}
\end{equation}

We study now the behavior of the renormalized density correlation function $C(r)$, Eq.~\eqref{eq:C-x-definition}, considering first the critical point, $G_0 \equiv G(l_0) = G_c$. Performing the RG transformation up to the scale $r$ and then evaluating the Gaussian integral as in Eq.~\eqref{eq:CqGaussian}, we obtain
\begin{equation}
C(r) \sim - g(r)Z^2(r)r^{-d-1} \sim - G_c r^{-2d}(r/l_0)^\zeta.
\end{equation}
According to Eq.~\eqref{eq:C-x-definition}, $C(\boldsymbol{x}-\boldsymbol{x}^\prime)$ at $\boldsymbol{x}\ne\boldsymbol{x}^\prime$ is a correlation function of Noether currents $\hat{{\cal J}_t}=(-i/2)(\hat{U}^\dagger\partial_{t}\hat{U}-\hat{U} \partial_t\hat{U}^\dagger)$ at $\boldsymbol{x}$ and $\boldsymbol{x}^\prime$.
It is known that surface integrals of conserved currents are symmetry generators \cite{DiFrancesco1997} and thus have scaling dimension zero at criticality \cite{Gross-LesHouches76,Wen1992scaling}. Thus, our currents $\hat{\mathcal{J}}_t$ should have scaling dimension $d$, implying that $C(\boldsymbol{x}-\boldsymbol{x'}) \propto |\boldsymbol{x}-\boldsymbol{x'}|^{-2d}$ and
hence $\zeta=0$.

Extending this analysis to $G_0$ around $G_c$, we get
\begin{equation}
\label{eq:Cr-scaling}
C(r)\simeq-\frac{G(r)}{r^{2d}} = -\frac{G_{c}}{r^{2d}}f\left(\frac{r}{l_{\text{corr}}}\right),
\end{equation}
where $f(x)$ is a universal scaling function with two branches describing both sides of the transition. The $x \ll 1$ behavior of $f(x)$ describes the critical point and its vicinity. According to Eq. \eqref{eq:G_Z_critical-scaling}, we have:
\begin{equation}
f(x\ll 1)\approx\begin{cases}
1+x^{1/\nu}, & G_{0}>G_{c} \,,\\
1-x^{1/\nu}, & G_{0}<G_{c}\,.
\end{cases}
\end{equation}
The $x \gg 1$ behavior of $f(x)$ corresponds to the ``metallic'' and ``insulating'' phases. On the ``metallic'' side of the transition, Eq.~\eqref{eq:Cr-scaling} should reproduce the asymptotics $C(r) \propto -1 / r^{d+1}$. In the area-law phase, one expects $C(r)$ to decay exponentially fast at infinity. To match this asymptotic behavior, the scaling function should satisfy:
\begin{equation}
f(x\gg 1)\sim\begin{cases}
x^{d-1}, & G_{0}>G_{c}\,,\\
e^{-x}, & G_{0}<G_{c}\,.
\end{cases}
\label{eq:Cr-scaling-large-x}
\end{equation}

It is worth emphasizing that the derivation of the NLSM and its RG analysis are under full control for $G\gg 1$. Thus, the analytical treatment of the transition is parametrically controllable at $\epsilon \ll 1$, when $G_c \gg 1$. At the same time, physically more interesting dimensions $d$ are integer, such as $d=2$ or $d=3$. On general grounds, we expect that the above scaling analysis also applies to $d=2$, 3, in similarity to $\epsilon$ expansions for Anderson transitions and statistical-mechanics transitions. Below, we study numerically the $d=2$ model to verify these scaling predictions and to estimate the critical exponent $\nu$.

\textbf{\textit{Particle number covariance and mutual information}} ---
According to the above analysis, ${\cal C}_{A}^{(2)}$ and  ${\cal S}_{E}$ scale as $\ell^{d-1}\ln \ell$ in the ``metallic'' phase and exhibit the area-law scaling both in the ``insulating'' phase and at criticality. In order to locate the transition and to probe the scaling around it, it is advantageous to consider an observable with three distinct forms of scaling, in analogy with the conductance around the Anderson transition. With this motivation, we
analyze the particle-number covariance of two regions, $A$ and $B$,
\begin{equation}
\overline{G_{AB}}=- \overline{\left\langle \left\langle \hat{N}_{A}\hat{N}_{B}\right\rangle \right\rangle} =-\int_{A}d^{d}\boldsymbol{x}\int_{B}d^{d}\boldsymbol{y}\,C(\boldsymbol{x}-\boldsymbol{y}).
\label{eq:GAB-definition}
\end{equation}
If both regions have characteristic size $\ell_A, \ell_B \sim \ell$, and the separation between those regions is also of the same order, the scaling of  $G_{AB}$  should read, according to 
Eqs.~\eqref{eq:Cr-scaling}--\eqref{eq:Cr-scaling-large-x},
\begin{equation}
\label{eq:Covariance:scaling}
\overline{G_{AB}}\simeq\widetilde{f}(\ell / l_\text{corr})\sim \begin{cases}
c_1 (\ell/l_{\text{corr}})^{d-1}, & G_{0}>G_{c}\,,\\
c_2 G_{c}, & G_{0}=G_{c}\,,\\
\exp(-c_3 \ell/l_{\text{corr}}), & G_{0}<G_{c} \,,
\end{cases}
\end{equation}
with numerical factors $c_{1,2,3} = O(1)$ depending on the exact geometry under consideration.
Since $C(\boldsymbol{x}-\boldsymbol{y})$ is a correlation function of Noether currents, $- C(\boldsymbol{x}-\boldsymbol{y})$ has a meaning of two-point conductance within the Anderson-transition analogy, and $G_{AB}$ has the meaning of conductance of a system with two leads attached to regions $A$ and $B$. Furthermore, $G_{AB}$ can be related to the mutual information $I(A:B)$  between these regions,
\begin{equation}
I(A:B)={\cal S}_{E}(A)+{\cal S}_{E}(B)-{\cal S}_{E}(A\cup B)\simeq\frac{2\pi^{2}}{3}G_{AB},
\end{equation}
implying that $I(A:B)$ exhibits the same scaling \eqref{eq:Covariance:scaling}. Thus, it can also be used as a  probe of the transition.

\begin{figure}[t]
    \centering
    \includegraphics[width=\columnwidth]{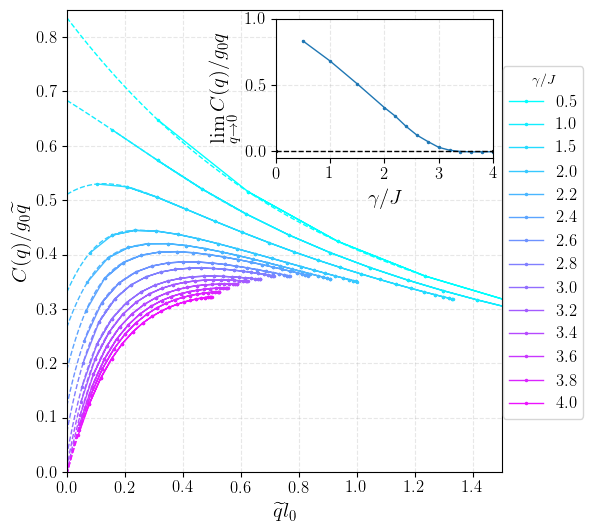}
    \caption{Correlation function $C(q)$ of an $L\times L$ system with $L=40$ across the transition. Shown are dependencies of $C(q) / g_0 \widetilde{q}$ on $\widetilde{q}l_0$ [with $\widetilde{q} = 2 \sin (q/2)$] for values of the measurement rate $\gamma / J$ from 0.5 to 4.0. Dashed lines: cubic polynomial extrapolation down to $q \to 0$ using 5 lowest momenta. Inset: the extrapolated value at $q \to 0$ as a function of $\gamma / J$. For $\gamma > \gamma_c$, where  $\gamma_c / J \simeq 3.2$, the extrapolated value is zero within numerical precision, which is a manifestation of the measurement-induced phase transition.}
    \label{fig:Cq}
\end{figure}

\textbf{\textit{Numerical analysis}} --- 
To verify and complement our analytical theory, we carried out direct numerical simulations of dynamics of a monitored two-dimensional free-fermion system following the protocol outlined above. As the protocol preserves the Gaussian nature of the state, the system is fully characterized by the single-particle Green function $\mathcal{G}_{\boldsymbol{x}\boldsymbol{x}^{\prime}}\equiv\left\langle \hat{\psi}^{\dagger}(\boldsymbol{x})\hat{\psi}(\boldsymbol{x}^{\prime})\right\rangle$. Executing unitary evolution along with projective measurements until the system has reached the steady state, we evaluated the two-point density correlation function on a lattice:
\begin{equation}    C_{\boldsymbol{x}\boldsymbol{x}^{\prime}}
\!=\!\left\langle \hat{n}(\boldsymbol{x})\hat{n}(\boldsymbol{x}^{\prime})\right\rangle -\left\langle \hat{n}(\boldsymbol{x})\right\rangle \left\langle \hat{n}(\boldsymbol{x}^{\prime})\right\rangle\!=\!\mathcal{G}_{\boldsymbol{x}\boldsymbol{x}}\delta_{\boldsymbol{x}\boldsymbol{x}^{\prime}}\!-\mathcal{G}_{\boldsymbol{x}\boldsymbol{x}^{\prime}}\mathcal{G}_{\boldsymbol{x}^{\prime}\boldsymbol{x}}.
\end{equation}
The protocol was repeated $\sim 100$ times, and averaging over quantum trajectories and over positions in the sample was performed to calculate $C(\boldsymbol{x}-\boldsymbol{x}^{\prime})=\overline{C_{\boldsymbol{x}\boldsymbol{x}^{\prime}}}$. 

The results obtained for an $L\times L$ system with periodic boundary conditions and $L=40$ are shown in Fig.~\ref{fig:Cq}. According to the above analytical discussion, the transition can be characterized by the behavior of $\lim\limits_{q\to0}C(q)/q$, which is expected to be finite in the ``metallic'' phase and zero in the ``insulating'' phase [where $C(q)\propto q^2$ due to exponential decay of $C(r)$]. The $q\to0$ values extrapolated from the finite-size curves (see inset) are fully consistent with this prediction, yielding an estimate $\gamma_c/J\approx3.2$ for the critical point. Furthermore, at $\gamma\to0$, the Gaussian approximation \eqref{eq:CqGaussian} is parametrically justified and predicts that $C(q)/g_0 q$ should saturate at unity, which is also in excellent agreement with the numerical data. For more details of numerical analysis of $C(q)$, see \cite{SuppMat}.

\begin{figure}[t]
\centering
\includegraphics[width=\columnwidth]{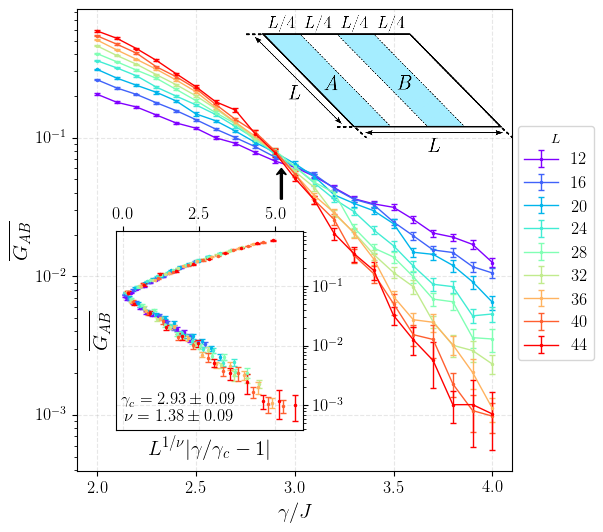}
\caption{Particle number covariance $\overline{G_{AB}}$ between two regions $A$ and $B$ in a geometrical setup shown in upper inset, as a function of ratio $\gamma / J$ for different system sizes $L$. Arrow marks the estimated position of the transition, manifesting itself as a crossing point of all curves, see Eq. \eqref{eq:Covariance:scaling}. Inset: collapse according to Eq.~\eqref{eq:Covariance:scaling} with a scaling function $\widetilde{f}(x^{\nu})$, where $x$ is given by the expression labeling the $x$-axis of the inset. Obtained values of $\gamma_c$ and $\nu$ are shown.}
\label{fig:covariance}
\end{figure}

To explore the transition more accurately, we focus now on the scaling of the particle number covariance \eqref{eq:GAB-definition} in a narrower vicinity of the transition point $\gamma_c$. We use a geometrical setup consisting of two regions of size $L/4 \times L$ separated by distance $L/4$, see upper inset of Fig. \ref{fig:covariance}.
According to Eqs. \eqref{eq:l-corr}, \eqref{eq:Covariance:scaling}, the data for $G_{AB}$ should collapse, with a suitable choice of $\gamma_c$ and critical exponent $\nu$, on a single universal function $\widetilde{f}(x)$. Figure  \ref{fig:covariance} demonstrates the behavior of $\overline{G_{AB}}$ as a function of $\gamma$ for system sizes from $L=12$ to $L=44$. A well-pronounced crossing point of all curves is observed, providing a more accurate value for the transition point, $\gamma_c / J \approx 2.9 \pm 0.1$. As shown in the lower inset, a good scaling collapse can be achieved, yielding the critical exponent $\nu = 1.4 \pm 0.1$. 
More general collapse of the form $\overline{G}_{AB} / L^\zeta = \widetilde{f}(L / \ell_{\text{corr}})$, which keeps the value of the critical exponent $\zeta$ finite, yields an indistinguishable from zero value $\zeta \sim 10^{-4}$, confirming our analytical prediction $\zeta = 0$. Collapse analysis was performed using \texttt{pyfssa} package \cite{PyFSSA, Melchert2009, SuppMat}, which yields $\gamma_c$ and $\nu$ with error bars.

\textbf{\textit{Summary}}---We have developed a field theory for  monitored  $d$-dimensional free-fermion systems, which has a form of an SU($R$) NLSM in $d+1$ dimensions, with the $R\to1$ replica limit. By performing its RG analysis, we have demonstrated the existence of a measurement-induced entanglement phase transition in such models for dimensions $d>1$. The transition point separates a symmetry-broken ``metallic'' phase realized for rare measurements and  characterized by $\ell^{d-1}\ln\ell$ scaling of the second particle-number cumulant and the entanglement entropy from the ``insulating'' area-law phase for frequent measurements. 
We have further calculated critical indices in the one-loop order of $\epsilon$-expansion in $d=1+\epsilon$ dimensions. Additionally, numerical investigation of the $d=2$ model on a square lattice enabled us to pinpoint the critical point and to estimate the correlation-length exponent $\nu = 1.4 \pm 0.1$. The measurement-induced entanglement transition for free fermions in dimensions $d>1$ bears a striking similarity to Anderson metal-insulator transition in disordered systems of $d+1$ dimensions. In the present context, it is the quantum information (closely related to particle-number fluctuations) that experiences localization.  As future important generalizations of the theory, we envisage inclusion of interaction and/or of static disorder, as well as measurement protocols generating non-trivial topology.

\textbf{\textit{Acknowledgments}}---We are grateful to A. Altland, D. Bernard, M. Buchhold, K. Chahine, S. Diehl, E. Doggen, Y. Gefen, I. Gruzberg, P. Ostrovsky, P. P\"opperl, M. Szyniszewski, and R. Vasseur for fruitful discussions. We acknowledge support by the Deutsche Forschungsgemeinschaft (DFG) via the grants MI {658/14-1} and GO {1405/6-1}.

\textit{Note added}---We have learned about a parallel related activity by K. Chahine and M. Buchhold \cite{chahine2023entanglement}. After submission of our paper, a preprint \cite{Jin2023} appeared, which reports a transition for a $d > 2$ \textit{single-particle} continuously monitored problem.

\widetext
\clearpage

\renewcommand{\theequation}{S\arabic{equation}}
\renewcommand{\thefigure}{S\arabic{figure}}
\renewcommand{\thesection}{S\arabic{section}}
\setcounter{equation}{0}
\setcounter{figure}{0}

\begin{center}
\Large \textbf{Supplemental Material to ``Measurement-induced phase transition for free fermions above one dimension''}
\end{center}

In this Supplemental Material, we provide technical details of our analytical formalism  as well as additional numerical results for the $d=2$ model studied in the main text. 

\section{Derivation of the non-linear sigma model (NLSM)}

In this Section we outline key steps of the derivation of the non-linear sigma model (NLSM) for the measurement problem from Ref. \cite{Poboiko2023}.

\subsection{Density matrix and replica trick}

 For each ``measurement trajectory'' $\mathcal{T} = \{t_m, \boldsymbol{x}_m, n_m\}_{m=1}^M$ with $M$ measurements at times $t_m$, locations $x_m$, and outcomes $n_m$, we define a non-normalized wavefunction:
\begin{equation}
\left|\widetilde{\Psi}({\cal T})\right>=\hat{U}_{0}(T,t_{M})\hat{\mathbb{P}}_{n_{M}}(\boldsymbol{x}_{M},t_{M})\hat{U}_{0}(t_{M},t_{M-1})\hat{\mathbb{P}}_{n_{M-1}}(\boldsymbol{x}_{M-1},t_{M-1})\dots\mathbb{P}_{n_1}(\boldsymbol{x}_1,t_1)\hat{U}_{0}(t_{1},t_{0})\left|\Psi_{0}\right>,
\end{equation}
with $\hat{U}_0(t_2,t_1) = \exp[-i\hat{H}(t_2-t_1)]$ being the  unitary evolution operator governed by the Hamiltonian $\hat{H}$. Equivalently, one can introduce a non-normalized density matrix associated with this wavefunction as $\hat{D}({\cal T})=\left|\widetilde{\Psi}({\cal T})\right>\left<\widetilde{\Psi}({\cal T})\right|$, so that the normalized density matrix is given by 
\begin{equation}
    \hat{\rho}({\cal T})=\hat{D}({\cal T})/{\rm Tr}\hat{D}({\cal T}).
\end{equation}
The norm of this wavefunction has a useful property that it provides a generalization of Born's rule for a set of consequent measurements, i.e., it yields a probability of given outcomes $\{n_m\}$ at fixed $\{t_m, \boldsymbol{x}_m\}$:
\begin{equation}
\label{eq:BornRule}
\text{Prob}({\cal T})=\left\langle \widetilde{\Psi}({\cal T})|\widetilde{\Psi}({\cal T})\right\rangle = \Tr \hat{D}(\mathcal{T}).
\end{equation}

In order to calculate the $N$th cumulant of an arbitrary observable (e.g., of the particle number in a subsystem) averaged over all measurement trajectories, one needs to average $N$ replicas of the density matrix $\hat{\rho} = \hat{D} / \Tr \hat{D}$ (we omit the argument $\mathcal{T}$ for brevity). Such average can be rewritten explicitly via matrix $\hat{D}$ as:
\begin{equation}
\label{eq:ReplicaTrick}
\overline{\hat{\rho}^{\otimes N}}^{({\cal T})}=\overline{\sum_{\{n_{m}\}}\Tr^{1-N}\hat{D}\cdot\hat{D}^{\otimes N}}^{(t_{m},\boldsymbol{x}_{m})} = \lim_{R\to1}{\rm Tr}_{r=N+1,\dots,R}\overline{\sum_{\{n_{m}\}}\prod_{r=1}^{R}\hat{D}^{(r)}}^{(t_{m},\boldsymbol{x}_{m})},
\end{equation}
where an additional factor $\Tr \hat{D}$ (hence, the trace is raised to power $1-N$) arises from the Born's rule \eqref{eq:BornRule}. In the last identity in  Eq.~\ref{eq:ReplicaTrick},
we have utilized a replica trick, which is required to get rid of the denominator that appears in the intermediate expression in Eq.~\eqref{eq:ReplicaTrick} for $N > 1$. Specifically, we have introduced $R \geq N$ copies of the matrix $\hat{D}$. Tracing out $R-N$ of these copies generates a prefactor $\Tr^{R-N}\hat{D}$ that yields the desired factor $\Tr^{1-N}\hat{D}$ after the analytic continuation $R \to 1$. Thus, it is Born's rule that is responsible for the unusual replica limit $R \to 1$ in the measurement problem.

The summation over measurement outcomes together with averaging over the independent and uniformly distributed measurement positions and Poissonian distributed measurement times can be performed conveniently within the Keldysh formalism. Introducing the Keldysh time contour ordering $\mathcal{T}_{\mathcal{C}}$, we obtain:
\begin{equation}
    \overline{\sum_{\{n_{m}\}}\prod_{r=1}^{R}\hat{D}^{(r)}}^{(t_{m},\boldsymbol{x}_{m})}={\cal T}_{{\cal C}}\left\{ \left(\prod_{r=1}^{R}\hat{\rho}^{(r)}_{0}\hat{U}^{(r)}_{\mathcal{C}}\right)\exp\left[\gamma\int\! d^{d+1}\boldsymbol{r}\,\left(\sum_{n_{m}=0,1}\prod_{r=1}^{R}\hat{\mathbb{P}}_{n_{m}}^{+,(r)}(\boldsymbol{x}_{m},t_{m})\hat{\mathbb{P}}_{n_{m}}^{-,(r)}(\boldsymbol{x}_{m},t_{m})-1\right)\right]\right\}.
\end{equation}
Here $\hat{\rho}_{0} = \left|\Psi_{0}\right>\left<\Psi_{0}\right|$ and $\hat{U}_{\mathcal{C}} = \hat{U}_{0}(T,t_0)\hat{U}_{0}(t_0,T)$ are the initial density matrix and the Keldysh contour evolution operator, correspondingly, superscript $(r)$ means that those operators act on a many-body Hilbert space of the $r$-th replica, and we use the short-hand notation $\int d^{d+1}\boldsymbol{r}\equiv\int_{t_{0}}^{T}dt\sum_{\boldsymbol{x}}$ for integration over time and summation over sites at which measurements were performed. As we are addressing the evolution of a density matrix, each measurement gives rise to two projection operators, one on the ``forward'' ($+$) and one on the ``backward'' ($-$) branches of the Keldysh contour ${\cal C}=(t_0,T)\cup(T,t_0)$.

\subsection{Keldysh fermionic path integral representation}
At the next step, we introduce a fermionic path integral representation in terms of Grassmanian fermionic fields $\psi_{r}^{\pm}(\boldsymbol{x},t)$ and $\bar{\psi}_{r}^{\pm}(\boldsymbol{x},t)$ that have indices corresponding to the Keldysh ($\pm$) and replica ($r = 1, \dots, R$) spaces. Utilizing the Larkin-Ovchinnikov rotation 
\begin{equation}
\begin{cases}
\psi_{1,2}=(\psi_{+}\pm\psi_{-})/\sqrt{2},\\
\bar{\psi}_{1,2}=(\bar{\psi}_{+}\mp\bar{\psi}_{-})/\sqrt{2},
\end{cases}
\label{eq:Keldysh-rotation}
\end{equation}
we obtain the fermionic action, which consists of two terms, 
\begin{equation}
{\cal L}[\psi,\bar{\psi}]={\cal L}_{0}[\psi,\bar{\psi}]+\gamma{\cal L}_{M}[\psi,\bar{\psi}].
\label{eq:action-two-terms}
\end{equation}
The first term describes the unitary evolution and contains information about the initial state:
\begin{equation}
{\cal L}_{0}[\psi,\bar{\psi}]=\sum_{r}\bar{\psi}_{r}\,\left[i\partial_{t}-\hat{H}+i\delta\begin{pmatrix}1 & 2 F_0\\
0 & -1
\end{pmatrix}_K\right]\,\psi_{r},
\end{equation}
with infinitesimal $\delta \to 0$ and with $F_0(\varepsilon)=2n_0(\varepsilon)-1$ being the Keldysh distribution function corresponding to the fermionic distribution function $n_0$ of the initial state at energy $\varepsilon$ (the subscript ``$K$'' refers to the Keldysh space). For brevity, we use here the same notation $\hat{H}$ for the single-particle tight-binding Hamiltonian that we used above for its second-quantized form. The second term in Eq.~\eqref{eq:action-two-terms}  describes a contribution to the fermionic Lagrangian, which arises from the projective measurements:
\begin{equation}
    {\cal L}_{M}[\psi,\bar{\psi}]=\prod_{r=1}^{R}P_{0}[\psi_{r},\bar{\psi}_{r}]+\prod_{r=1}^{R}P_{1}[\psi_{r},\bar{\psi}_{r}]-1\,,
\end{equation}
where
\begin{equation}
    P_{0}[\psi,\bar{\psi}]=1-\bar{\psi}_{1}\psi_{2}-\bar{\psi}_{2}\psi_{1}-\bar{\psi}_{1}\psi_{1}\bar{\psi}_{2}\psi_{2}\,,\qquad P_{1}[\psi,\bar{\psi}]=-\bar{\psi}_{1}\psi_{1}\bar{\psi}_{2}\psi_{2}.
\end{equation}

\subsection{Non-linear sigma model}
To proceed with the derivation of the NLSM field theory, we introduce a $Q$-matrix of size $2R \times 2R$ as a bilinear combination of fermionic fields 
\begin{equation}
Q_{rr^{\prime}}^{\tau\tau^{\prime}}(\boldsymbol{x}, t)=2\left\langle \psi_{r}^{\tau}(\boldsymbol{x}, t)\bar{\psi}_{r^{\prime}}^{\tau^{\prime}}(\boldsymbol{x},t)\right\rangle,
\end{equation}
with $r,r^\prime = 1,\dots,R$ being replica indices and $\tau = 1,2$ being Keldysh indices corresponding to the Larkin-Ovchinnikov transformation. This allows us to transform a functional integral over the fermionic fields into that over the $Q$-matrix field.
For this purpose, we utilize a generalized Hubbard-Stratonovich transformation and carefully regularize terms in the diagrammatic expansion, which include fermionic loops at exactly coinciding coordinates, see Ref.~\cite{Poboiko2023} for technical details.
This leads to an effective action that describes the large-distance and long-time asymptotic behavior of the problem, which again consists of two terms:
\begin{equation}
\label{eq:NLSM:S}
S[\hat{Q}] = S_0[\hat{Q}] + \gamma \int d^{d+1} \boldsymbol{r} \mathcal{L}_M[\hat{Q}].
\end{equation}
Here,
\begin{equation}
\label{eq:NLSM:S0}
i S_0[\hat{Q}] = \Tr\left(\ln(i\partial_{t}-\hat{H}_{0}+i\gamma\hat{Q})-\frac{\gamma}{2}\hat{Q}^{2}\right) \approx {\rm Tr}\left(\frac{1}{2}\hat{\Lambda}\hat{{\cal R}}^{-1}\partial_{t}\hat{{\cal R}}-\frac{D}{8}(\nabla\hat{Q})^{2}\right),
\end{equation}
describes the ``own'' dynamics of the system and
\begin{equation}
\label{eq:NLSM:SM}
{\cal L}_{M}[\hat{Q}]=\det\left(\frac{1-\hat{Q}\hat{\tau}_{x}}{2}\right)+\det\left(\frac{1+\hat{Q}\hat{\tau}_{x}}{2}\right)-1 
\end{equation} 
is the measurement-induced Lagrangian.
The $Q$ field is subjected to the conventional NLSM  constraint  $\hat{Q}^2 = 1$ together with $\Tr \hat{Q} = 0$, and the integration manifold is spanned by rotations of the self-consistent Born approximation (SCBA) saddle point $\hat{\Lambda}$ as:
\begin{equation}
    \hat{Q}=\hat{{\cal R}}\hat{\Lambda}\hat{{\cal R}}^{-1} \in \mathrm{U}(2R) / \mathrm{U}(R) \times \mathrm{U}(R),\quad\hat{\Lambda}=\int_{BZ}\frac{d^{d}\boldsymbol{k}}{(2\pi)^{d}}\begin{pmatrix}1 & 2F_{0}(\varepsilon_{\boldsymbol{k}})\\
0 & -1
\end{pmatrix}_{K}=\begin{pmatrix}1 & 2(1-2\rho)\\
0 & -1
\end{pmatrix}_{K}.
\end{equation}
The SCBA describes an infinite-temperature state with the band filling factor $\rho \in [0,1]$ (which is inherited from the initial state, as the protocol conserves the total number of particles), with the decay rate $1/2\tau = \gamma$ of the single-particle plane-wave excitations. Note that the density $\rho$ is the only information about the initial state, which is left in the theory in the limit of long times over which the measurements are performed.  
The diffusion constant $D$, which arises after performing the standard gradient expansion [see the last expression in Eq.~\eqref{eq:NLSM:S0}], describes the long-wavelength fluctuations of density. It is related to the root-mean-square group velocity $v_0$ of excitations in our tight-binding model as
\begin{equation}
    D=v^{2}\tau/d = J^2/\gamma,\quad v^{2}=\int_{BZ}\frac{d^{d}\boldsymbol{k}}{(2\pi)^{d}}\left(\frac{\partial\varepsilon_{\boldsymbol{k}}}{\partial\boldsymbol{k}}\right)^{2}=2dJ^{2}\,.
\end{equation}

An analysis of symmetries of the action given by Eqs.~(\ref{eq:NLSM:S}-\ref{eq:NLSM:SM}) reveals that it is useful to parametrize the integration manifold as follows:
\begin{equation}
\hat{Q}=\hat{{\cal R}}_{\Phi}\hat{{\cal R}}_{\Theta}\hat{Q}_{0}\hat{{\cal R}}_{\Theta}^{-1}\hat{{\cal R}}_{\Phi}^{-1},\quad\hat{{\cal R}}_{\Phi}=e^{i\hat{\Phi}\hat{\tau}_{x}/2},\quad\hat{{\cal R}}_{\Theta}=e^{i\hat{\Theta}\hat{\tau}_{y}/2},\quad \Tr \hat\Phi = \Tr \hat\Theta = 0 \,.
\end{equation}
Here, the two replica-symmetric generators responsible for the fluctuations of the average (non-postselected) density are combined in the replica-symmetric Goldstone field $\hat{Q}_0 \in \mathrm{U}(2) / \mathrm{U}(1) \times \mathrm{U}(1) = S^2$. Further, $(R^2-1)$ replica-asymmetric generators that commute with the matrix $\hat{\tau}_x$ in the Keldysh space [cf. Eq.~\eqref{eq:NLSM:SM}] yield another set of Goldstone modes that span the manifold $\hat{U} = \exp(i \hat{\Phi}) \in \mathrm{SU}(R)$. Finally, the remaining $(R^2 - 1)$ generators denoted as $\hat{\Theta}$ describe massive modes that can be safely integrated out in the Gaussian approximation. 

The theory now splits into two sectors.
The replica-symmetric sector that corresponds to fluctuations of the density averaged over measurements outcomes is described by an action of the same form \eqref{eq:NLSM:S0} but with the matrix $\hat{Q}$ (and $\hat{\mathcal{R}}$) replaced with the replica-symmetric mode $\hat{Q}_0$ (and replica-symmetric generators $\hat{\mathcal{R}_0}$ correspondingly). This implies diffusion for the averaged density, with the diffusion constant $D$ (which is not subjected to any renormalization~\cite{Poboiko2023}). 

For the calculation of the particle-number fluctuations, we are interested in the replica-asymmetric (``replicon'') sector that is described by the $(d+1)$-dimensional Euclidian action with the integration weight $\exp\left(-\int d^{d+1}\boldsymbol{r}{\cal L}[\hat{U},\hat{Q}_0]\right)$:
\begin{equation}
\label{eq:NLSM:SUR}
    {\cal L}[\hat{U},\hat{Q}_0]=\frac{g[\hat{Q}_0]}{2}\Tr\left(\frac{1}{v_{0}}\partial_{t}\hat{U}\partial_{t}\hat{U}^{\dagger}+v_{0}\nabla\hat{U}\nabla\hat{U}^{\dagger}\right),
\end{equation}
with $v_0 = v / \sqrt{d}$. The ``coupling constant'' $g$ here is given by:
\begin{equation}
g[\hat{Q}_0] = 2 \sqrt{D \tau} \rho[\hat{Q}_0] (1-\rho[\hat{Q}_0]) \approx 2\rho(1-\rho) l_0/\sqrt{d},
\end{equation}
with $l_0 = v \tau$ and $\rho[\hat{Q}_0] = \frac{1}{4}\Tr(1-\hat{Q}_{0}\hat{\tau}_{x})$ being average (replica-symmetric) fermionic density.
Since $\hat{Q}_0$ exhibits fluctuations of the replica-symmetric sector, the dependence of $g$ on $\hat{Q}_0$ leads to a coupling between the replica-symmetric and replicon sectors. However, this coupling leads only to small, infrared-finite corrections and can be safely neglected, which corresponds to replacement $\rho[\hat{Q}_0] \to \rho[\hat{\Lambda}] = \rho$. The final form of action in the replicon sector is 
\begin{equation}
\label{eq:NLSM:SUR-2}
    {\cal L}[\hat{U}]=\frac{g}{2}\Tr\left(\frac{1}{v_{0}}\partial_{t}\hat{U}\partial_{t}\hat{U}^{\dagger}+v_{0}\nabla\hat{U}\nabla\hat{U}^{\dagger}\right),
\end{equation}
with the bare value of the coupling $g$ being $g_0=\rho(1-\rho)v_0/\gamma$. Thus, we arrive at Eq.~(7) 
(in the absence of source field) and Eq.~(8) 
of the main text.

The time integration in the action for $\hat{U}$ modes is performed over times $t \in [t_0, T]$. As we are interested in the steady state, we take limit $t_i \to -\infty$. The time $T$ corresponds to the moment at which we study the correlation function, and, since the action \eqref{eq:NLSM:SUR} involves the second derivative with respect to time, it has to be supplied with a boundary condition at $t=T$. The correct boundary conditions can be obtained by noting that if one continues the Keldysh contour up to $t = +\infty$ but effectively puts $\gamma = 0$ at $t \in (T, +\infty)$, the forward- and backward-time evolution at the added part will be purely unitary and will thus cancel. On the other hand, on the level of NLSM, putting $\gamma = 0$ corresponds to the coupling constant $g \to \infty$, which ``freezes'' the value $\hat{U}(x, t > T) = \const = \hat{\mathbb{I}}$, thus determining the boundary condition at $t=T$ stated in the main text below Eq.~(7).  The time $T$ (at which we calculate observables of interest) can be chosen to be $T = 0$ without loss of generality. 

\subsection{Sources, generating functional, and density correlation function}

Since we study the density-correlation function, we should express the density operator via the fields described by the NLSM. To do this, we note that the ``classical'' Keldysh component of the fermionic density in replica $r$ is given by a bilinear combination of fermionic fields, i.e., is linear in the $\hat{Q}$-matrix:
\begin{equation}
\rho_{r}(\boldsymbol{x},t)=\frac{1}{4}\Tr_K(1-\hat{Q}_{rr}\hat{\tau}_{x}).
\end{equation}
Thus, we can introduce a generating functional with a source field 
 $\hat{\xi} = \mathrm{diag}(\{\xi_r\}_{r=1}^{R})$ 
 (diagonal in the replica space) as 
\begin{equation}
{\cal Z}[\hat{\xi}]=\left\langle \exp\left(i\int d^{d+1}\boldsymbol{r}\sum_{r}\xi_{r}(\boldsymbol{r})\rho_{r}(\boldsymbol{r})\right)\right\rangle =\int{\cal D}\hat{Q}\exp\left[iS[\hat{Q}]+\frac{i}{4}\Tr\big(\hat{\xi}(1-\hat{Q}\hat{\tau}_{x})\big)\right].
\end{equation}
As we proceed with the derivation as outlined above, the density source then naturally splits into two contributions: the replica-symmetric source $\xi_0 = \Tr_R(\hat{\xi}) / R$ that couples to the replica-symmetric modes $\hat{Q}_0$, and the ``replicon'' source, $\hat{\Xi} = \hat{\xi} - \xi_0$, which couples to $\hat{U}$. This leads to the generating functional $\mathcal{Z}[\hat{\Xi}]$ in the replicon sector given by an NLSM path integral over the SU$(R)$ matrix fields $\hat{U}$ [here, $\boldsymbol{r} = (\boldsymbol{x}, t)$ and the time integration is constrained to $t < 0$]:
\begin{equation}
\mathcal{Z}[\hat{\Xi}]=\int{\cal D}\hat{U}{\cal D}\hat{U}^{\dagger}\exp\left(-\int d^{d+1}\boldsymbol{r}~{\cal L}[\hat{U}(\boldsymbol{r}),\hat{\Xi}(\boldsymbol{r})]\right),
\end{equation}
The action for this generating functional is given by Eq.~(7) 
of the main text, with the source field entering via the ``long derivative'', Eq.~(9) 
of the main text. 

In the replicated field theory, the density-correlation function becomes a matrix in the replica space: $C_{ab}(\boldsymbol{r},\boldsymbol{r}^{\prime})=\left\langle \rho_{a}(\boldsymbol{r})\rho_{b}(\boldsymbol{r}^{\prime})\right\rangle $. The diagonal elements of this matrix correspond to the correlation function $C_0$ governed by the averaged density matrix:
\begin{equation}
C_{a=b}(\boldsymbol{r},\boldsymbol{r}^{\prime})=\overline{\left\langle \{\hat{n}(\boldsymbol{r}),\hat{n}(\boldsymbol{r}^{\prime})\}/2\right\rangle }-\rho^{2}=C_{0}(\boldsymbol{r},\boldsymbol{r}^{\prime}).
\end{equation}
On the other hand, off-diagonal elements at $t = t^\prime = t_f \equiv 0$ correspond to
\begin{equation}
C_{a\neq b}(\boldsymbol{x},\boldsymbol{x}^{\prime},t=0)\equiv\overline{\left\langle \hat{n}(\boldsymbol{x},t=0)\right\rangle \left\langle \hat{n}(\boldsymbol{x}^{\prime},t=0)\right\rangle }-\rho^{2}\equiv C_{0}(\boldsymbol{r},\boldsymbol{r}^{\prime})-C(\boldsymbol{x},\boldsymbol{x}^{\prime},t=0).
\end{equation}
Thus, the correlation function of our interest, $C(\boldsymbol{x}, \boldsymbol{x}^\prime)$, is given by a difference between the diagonal and off-diagonal elements of matrix $C_{ab}$:
\begin{equation}
C(\boldsymbol{x},\boldsymbol{x}^{\prime})\equiv\overline{\left\langle \hat{n}(\boldsymbol{x})\hat{n}(\boldsymbol{x}^{\prime})\right\rangle }-\overline{\left\langle \hat{n}(\boldsymbol{x})\right\rangle \left\langle \hat{n}(\boldsymbol{x}^{\prime})\right\rangle }=C_{a=b}(\boldsymbol{x},\boldsymbol{x}^{\prime})-C_{a\neq b}(\boldsymbol{x},\boldsymbol{x}^{\prime})=\frac{1}{R-1}\left(\sum_{a}C_{aa}(\boldsymbol{x},\boldsymbol{x}^{\prime})-\frac{1}{R}\sum_{ab}C_{ab}(\boldsymbol{x}, \boldsymbol{x}^\prime)\right).
\label{eq:C-Cab}
\end{equation}
Clearly, the replica-symmetric sector (which does not distinguish between $C_{a=b}$ and $C_{a \ne b}$) does not affect the r.h.s. of Eq.~\eqref{eq:C-Cab}. 
Therefore, only the replicon ($\hat{U}$) modes contribute to the correlation function of our interest, so that we can focus entirely on the replicon sector. We remind the reader that, to obtain physical observables, one should take the  $R\to 1$ limit of correlation functions derived within the replica formalism [e.g., in the r.h.s. of Eq.~\eqref{eq:C-Cab}].

To obtain the matrix correlation function $C_{ab}$, one needs to expand the logarithm of the generating functional to the second order in $\hat{\Xi}$:
\begin{equation}
C_{ab}(\boldsymbol{r}_{1},\boldsymbol{r}_{2})\equiv-\frac{\delta^{2}\ln\mathcal{Z}[\hat{\Xi}]}{\delta\Xi_{a}(\boldsymbol{r}_{1})\delta\Xi_{b}(\boldsymbol{r}_{2})}=\frac{g_{0}}{2v_{0}}\left(\left\langle U_{ab}(\boldsymbol{r}_{1})U_{ba}^{\dagger}(\boldsymbol{r}_{2})\right\rangle +\delta_{ab}\right)\delta(\boldsymbol{r}_{1}-\boldsymbol{r}_{2})-\frac{g_{0}^{2}}{v_{0}^{2}}\left\langle \mathcal{J}_{t,aa}(\boldsymbol{r}_1) \mathcal{J}_{t,bb}(\boldsymbol{r}_2)\right\rangle.
\label{eq:Cab-def}
\end{equation}
Here, we have introduced the $t$-component of the Noether current, $$\hat{{\cal J}}_t(\boldsymbol{r})=-\frac{i}{2}\left\{ \hat{U}^{\dagger}(\boldsymbol{r}),\partial_{t}\hat{U}(\boldsymbol{r})\right\},$$ and the average is calculated with respect to the same action without the source term. Further, at the boundary $t \to 0$, we have $\hat{U}_{ab}(\boldsymbol{r}) \to \delta_{ab}$, and thus the boundary correlation function simplifies:
\begin{equation}
C_{ab}(\boldsymbol{x}_{1},\boldsymbol{x}_{2})=\lim_{t_{1,2}\to0}\left[\frac{g_{0}}{v_{0}}\delta_{ab}\delta(\boldsymbol{x}_{1}-\boldsymbol{x}_{2})\delta(t_{1}-t_{2})-\frac{g_{0}^{2}}{v_{0}^{2}}\left\langle \mathcal{J}_{t,aa}(\boldsymbol{r}_1) \mathcal{J}_{t,bb}(\boldsymbol{r}_2)\right\rangle \right].
\label{eq:Cab-boundary}
\end{equation}
Furthermore, utilizing $\Tr\hat{\mathcal{J}}_t = 0$, we note that it satisfies following ``sum rule'':
\begin{equation}
\frac{1}{R}\sum_{ab}C_{ab}(\boldsymbol{r}_{1},\boldsymbol{r}_{2})=\frac{g_{0}}{v_{0}}\delta(\boldsymbol{r}_{1}-\boldsymbol{r}_{2}).
\end{equation}
Substituting it and the definition given by the first part of Eq. \eqref{eq:Cab-def} to Eq. \eqref{eq:C-Cab} yields Eq. (10) from the main text. Finally, combining it with Eq.~\eqref{eq:Cab-boundary} produces following equation for the physical density-correlation function:
\begin{equation}
\label{eq:C}
C(\boldsymbol{x}_{1},\boldsymbol{x}_{2})=\lim_{t_{1,2}\to0}\left[\frac{g_{0}}{v_{0}}\delta(\boldsymbol{x}_{1}-\boldsymbol{x}_{2})\delta(t_{1}-t_{2})-\frac{g_{0}^{2}}{v_{0}^{2}}\frac{1}{R-1}\sum_{a}\left\langle {\cal J}_{t,aa}(\boldsymbol{r}_{1}){\cal J}_{t,aa}(\boldsymbol{r}_{2})\right\rangle \right].
\end{equation}

\section{Correlation function in the Gaussian approximation}

Employing the exponential parametrization $\hat{U}(\boldsymbol{r})=e^{i\hat{\Phi}(\boldsymbol{r})}$ with a traceless matrix $\Phi$, one has ${\cal J}_{t,aa}(\boldsymbol{r})=\partial_{t}\Phi_{aa}(\boldsymbol{r})$ at the boundary. Within the Gaussian approximation (i.e., after expanding the action in small fluctuations of the field up to the second order), the Lagrangian for the $\Phi$ fields reads:
\begin{equation}
{\cal L}_{0}[\hat{\Phi}]\approx\frac{g}{2}\Tr\left[\frac{1}{v_{0}}(\partial_{t}\hat{\Phi})^{2}+v_{0}(\nabla\hat{\Phi})^{2}\right].
\end{equation}
Taking into account the boundary condition $\Phi(x,t=0) = 0$, the bare correlation function of $\Phi$ fields (i.e., that in the Gaussian approximation) is given by:
\begin{equation}
\label{eq:PhiGFGaussian}
\left\langle \Phi_{ab}(\boldsymbol{r}_{1})\Phi_{a^{\prime}b^{\prime}}(\boldsymbol{r}_{2})\right\rangle =\left(\delta_{ab^{\prime}}\delta_{a^{\prime}b}-\frac{1}{R}\delta_{ab}\delta_{a^{\prime}b^{\prime}}\right)\times\int_{0}^{\infty}\frac{d\omega}{\pi}\int\frac{d^{d}\boldsymbol{q}}{(2\pi)^{d}}e^{i\boldsymbol{q}(\boldsymbol{x}_{1}-\boldsymbol{x}_{2})}\sin\omega t_{1}\sin\omega t_{2}\frac{v_{0}/g_{0}}{\omega^{2}+v_{0}^{2}q^{2}}.
\end{equation}
Combining this with Eq.~\eqref{eq:C}, performing the Fourier transformation with respect to the coordinate difference $\boldsymbol{x}_1-\boldsymbol{x}_2$, and making use of the integral representation for the delta-function,
\begin{equation}
\delta(t_{1}-t_{2})=\int_{0}^{\infty}\frac{d\omega}{\pi}\cos\omega(t_{1}-t_{2}),
\end{equation}
we obtain:
\begin{equation}
\label{eq:CqGaussian-appendix}
C(\boldsymbol{q})=\frac{g_{0}}{v_{0}}\int_{0}^{\infty}\frac{d\omega}{\pi}\left[1-\frac{\omega^{2}}{\omega^{2}+v_{0}^{2}q^{2}}\right]=g_{0}|q|.
\end{equation}
The two terms in square brackets in Eq.~\eqref{eq:CqGaussian-appendix} correspond to the two terms in Eq.~\eqref{eq:C}, with the factor $\omega^2$ in the second term originating from time derivatives in currents $\mathcal{J}_t$, and integration over $\omega$ comes from the limit $t_{1,2} \to 0$.
Equation \eqref{eq:CqGaussian-appendix} is exactly the second part of Eq.~(11) 
of the main text. Finally, performing the inverse Fourier transformation in the spherical coordinates, we obtain:
\begin{equation}
C(\boldsymbol{x})=g_{0}\int\frac{d^{d}\boldsymbol{q}}{(2\pi)^{d}}|q|e^{i\boldsymbol{q}\boldsymbol{x}}=\frac{g_{0}}{(2\pi)^{d/2}}\frac{1}{x^{d/2-1}}\int_{0}^{\infty}q^{d/2+1}J_{d/2-1}(qx)dq=-\frac{\Gamma\left(\frac{d+1}{2}\right)}{\pi{}^{(d+1)/2}}\frac{g_{0}}{|\boldsymbol{x}|^{d+1}},
\label{eq:CxGaussian}
\end{equation}
where $\Gamma(z)$ is the gamma function and the integral over $q$ was evaluated utilizing the exponential regularization. This is the first part of Eq. (11) 
of the main text.

The Gaussian-approximation results \eqref{eq:CqGaussian-appendix}, \eqref{eq:CxGaussian} remain valid in the ``metallic'' phase $\gamma < \gamma_c$, up to a renormalization of the prefactor $g_0$ due to loop corrections (``weak localization''). The long-range (power-law) behavior \eqref{eq:CqGaussian-appendix}, \eqref{eq:CxGaussian} is a manifestation of the symmetry-broken (Goldstone) character of the phase for rare measurements ($\gamma < \gamma_c$). This behavior leads to a logarithmic $\ln\ell$ enhancement of the entanglement entropy (in comparison to the area-law phase). From this point of view, the scaling of the particle-number fluctuations and the entanglement entropy in the ``metallic'' phase may be termed ``critical'' (in analogy to, e.g., the ``critical'' scaling of entanglement in the ground state of a Fermi gas or in conformal fields theories).  We prefer, however, to avoid using the term ``critical'' regarding the $\gamma < \gamma_c$ phase, since this might lead to confusion with the critical point of the transition.  Instead, we use in the main text the term ``gapless'', which points to the existence of Goldstone modes (``diffusons'') in this symmetry-broken phase.

\section{One loop renormalization group equations in $d = 1 + \epsilon$}
In this Section we will extend the one-loop renormalization group calculation of Ref. \cite{Poboiko2023} to spatial dimensions $d = 1+\epsilon$ with $\epsilon \ll 1$. Let's consider action given by Eq. (7) of the main text, described by coupling constants $g$ and $Z$, and derive one-loop renormalization group equations for these coupling constants. Restricting the source $\hat{\Xi}$ to the boundary, and rescaling the time as $t \mapsto v_0 t$, the action acquires following form:
\begin{equation}
\mathcal{L}[\hat{U},\hat{\Xi}]=\frac{g}{2}\int d^{d+1}\boldsymbol{r}\Tr\left(\partial_{\mu}\hat{U}\partial_{\mu}\hat{U}^{\dagger}\right)+gZ\int d^d\boldsymbol{x}\Tr\left(\partial_{t}\hat{\Phi}(\boldsymbol{x},t=0)\hat{\Xi}(\boldsymbol{x})\right).
\end{equation}

We will use the background field method and decouple ``fast'' and ``slow'' modes of matrix $\hat{U}$ in a way that preserves unitarity as $\hat{U} = \hat{U}_{f} \hat{U}_{0}$. The interaction between fast and slow modes then reads:
\begin{equation}
{\cal L}_{\text{int}}[\hat{U}_f, \hat{U}_0]=-g\Tr\left[\hat{W}_{\mu}\partial_{\mu}\hat{U}_{0}\hat{U}_{0}^{\dagger}\right],\quad \hat{W}_{\mu}\equiv-i\hat{U}_{f}^{\dagger}\partial_{\mu}\hat{U}_{f} \approx \partial_{\mu}\hat{\Phi}_{f}-\frac{i}{2}\left[\hat{\Phi}_{f},\partial_{\mu}\hat{\Phi}_{f}\right]-\frac{1}{6}\left[\hat{\Phi}_{f}, \left[\hat{\Phi}_{f}, \partial_\mu \hat{\Phi}_{f}\right]\right],
\end{equation}
where in the last equation we have also used exponential parametrization for the ``fast'' modes $\hat{U}_f = \exp(i \hat{\Phi}_f)$.

\subsection{Renormalization of ``conductance'' $G$}
\label{sec:RG-G}

The renormalization of the ``bulk'' coupling constant $g$ then comes from the second order of perturbation theory in $\mathcal{L}_{\text{int}}$, which produces following correction to the effective action:
\begin{equation}
    S_{\text{eff}}^{(1)}[\hat{U}_0]=-\frac{g^{2}}{8}\int d^{d+1}\boldsymbol{r}_{1}d^{d+1}\boldsymbol{r}_{2}\Big\llangle \Tr\left(\left[\hat{\Phi}_{f},\partial_{\mu}\hat{\Phi}_{f}\right]\partial_{\mu}\hat{U}_{0}\hat{U}_{0}^{\dagger}\right)_{\boldsymbol{r}_{1}}\Tr\left(\left[\hat{\Phi}_{f},\partial_{\nu}\hat{\Phi}_{f}\right]\partial_{\nu}\hat{U}_{0}\hat{U}_{0}^{\dagger}\right)_{\boldsymbol{r}_{2}}\Big\rrangle,
\end{equation}
with angular brackets denoting average over ``fast'' fluctuations. Note that for the renormalization of the ``bulk'' coupling constant, the presence of the boundary is irrelevant and one can extend the time integration to the whole $(d+1)$-dimensional space. Performing the Wick contraction, and switching the integration to the ``center-of-mass'' $\boldsymbol{R} = (\boldsymbol{r}_1 + \boldsymbol{r}_2) / 2$ and relative motion $\boldsymbol{\rho} = \boldsymbol{r}_1 - \boldsymbol{r}_2$, we obtain that the correction renormalizes coupling constant $g$, and the correction reads:
\begin{equation}
\label{eq:RG-correction}
\delta g=-g^2 R\,\mathcal{I}_1,\quad \mathcal{I}_1=\frac{1}{d+1}\int d^{d+1}\boldsymbol{\rho}(\partial_{\mu}G_{f}(\boldsymbol{\rho}))^{2},
\end{equation}
with the bulk Green function of $\Phi_f$-fields in the momentum representation reading (cf. Eq. \eqref{eq:PhiGFGaussian}):
\begin{equation}
\label{eq:GfBulk}
G_{f}(\boldsymbol{q})=\frac{1}{g \boldsymbol{q}^2} \theta(\Lambda-|q|)\theta(|q|-\Lambda^{\prime}).
\end{equation}
In dimensions $d = 1+\epsilon$, the integral $\mathcal{I}$ then reads:
\begin{equation}
\mathcal{I}_1=\frac{1}{(2+\epsilon)g^{2}}\int_{\Lambda^\prime}^{\Lambda}\frac{d^{2+\epsilon}\boldsymbol{q}}{(2\pi)^{2+\epsilon}}\frac{1}{\boldsymbol{q}^{2}} \underset{\epsilon\ll1}{\approx}\frac{1}{g^{2}}\frac{\Lambda^{\epsilon}-\Lambda^{\prime\epsilon}}{4\pi\epsilon}.
\end{equation}
Introducing dimensionless ``conductance'' $G = g / \Lambda^\epsilon$, switching to real space cutoff $\ell = \Lambda^{-1}$, and taking the derivative with respect to $\ell$, the  correction \eqref{eq:RG-correction} then yields:
\begin{equation}
\label{eq:supp:betaG}
dG / d\ln \ell=\epsilon\, G-R/4\pi+O(1/G),
\end{equation}
with the first term coming from the explicit dependence of $G$ on $\Lambda$, and the second term coming from the renormalization of the ``conductivity'' $g$. Equation \eqref{eq:supp:betaG} is Eq. (13) of the main text. It fully agrees with Eq. (27) from Ref. \cite{Konig2012}, where $\epsilon=0$ and $\sigma = 4 \pi g$ (as can be seen from the form of the action given by Eq. (3) of Ref.~\cite{Konig2012}). The proportionality of the one-loop contribution to $(-R)$ also agrees with the $\beta$-function expansion given by Eq. (11) of Ref.~\cite{HikamiRG} and by Eq. (2.2) of Ref.~\cite{WegnerRG} [for group $\mathrm{SU}(N)$; presented also in Table III of review \cite{evers08}, in a line corresponding to the Hamiltonian class AIII]. We note that $\beta$-functions of Refs.~\cite{HikamiRG} and \cite{WegnerRG} were given for a different normalization of the coupling constant $t$ (proportional to our $G^{-1}$) which was not specified in these references.

\subsection{Renormalization of the density source $Z$}

The one loop renormalization of the ``boundary'' coupling constant $Z$ comes from the second order of perturbation theory, with one fast field put at the boundary and a single cubic interaction vertex:
\begin{equation}
    S^{(2)}_{\text{eff}}[\hat{U}_0,\hat{\Xi}]\approx-\frac{g^{2}Z}{6}\int d^{d+1}\boldsymbol{r}_{1}d^d\boldsymbol{x}_{2}\left\llangle \Tr\left(\left[\hat{\Phi}_{f}, \left[\hat{\Phi}_{f}, \partial_\mu \hat{\Phi}_{f}\right]\right](-i)\partial_{\mu}\hat{U}_{0}\hat{U}_{0}^{\dagger}\right)_{\boldsymbol{r}_{1}}\Tr\left(\partial_{t}\hat{\Phi}_{f}\hat{\Xi}\right)_{\boldsymbol{x}_{2}}\right\rrangle,
\end{equation}
where the time integration is taken explicitly over the semi-infinite space $t_1 < 0$. 
Similarly to Sec.~\ref{sec:RG-G}, we perform Wick contraction and switch to the ``center-of-mass'' $\boldsymbol{X} = (\boldsymbol{x}_1 + \boldsymbol{x}_2) / 2$ coordinate in real space and relative motion coordinate $\boldsymbol{\rho} = (\boldsymbol{y}=\boldsymbol{x}_1 - \boldsymbol{x}_2, t_1)$, in space and time domain. As a result, we find that this correction renormalizes the product $g Z$ in the original action, with the correction reading:
\begin{equation}
\delta(gZ)=-g^{2}ZR\,\mathcal{I}_{2},\quad\mathcal{I}_{2}=\int_{-\infty}^{0} dt_1 \int d^{d}\boldsymbol{y} \left[\partial_{t^{\prime}}G_{f}(0,t_1,t^{\prime})\partial_{t^{\prime\prime}}G(\boldsymbol{y},t_1,t^{\prime\prime})\right]_{t^{\prime\prime}\to0}^{t^{\prime}\to t_1} \,.
\end{equation}
Calculation of this integral with the Green functions given by
\begin{equation}
G_{f}(\boldsymbol{x},t_{1},t_{2})=\int_{0}^{\infty}\frac{d\omega}{\pi}\int_{\Lambda^{\prime}}^{\Lambda}\frac{d^{d}\boldsymbol{q}}{(2\pi)^{d}}e^{i\boldsymbol{q}\boldsymbol{x}}\sin\omega t_{1}\sin\omega t_{2}\frac{1}{g(\omega^{2}+\boldsymbol{q}^{2})}
\end{equation}
(cf. \eqref{eq:PhiGFGaussian} and note that it reduces to the Fourier transform of \eqref{eq:GfBulk} away from the boundary at $t_{1,2} \to -\infty$)
yields $\mathcal{I}_2 = \mathcal{I}_1$ to the leading order, i.e., the correction can be entirely ``absorbed'' in the renormalization of $g$. This result implies the absence of one-loop renormalization of $Z$, which is  Eq. (14) of the main text.

\section{Fluctuations of number of particles}

For an arbitrary region $A$ in the $d$-dimensional space, the second cumulant characterizing fluctuations of the number of particles is given by:
\begin{equation}
\label{eq:CumulantA}
C_{A}^{(2)}=\int_{A}d^{d}\boldsymbol{x}\int_{A}d^{d}\boldsymbol{y}\, C(\boldsymbol{x}-\boldsymbol{y})=\int\frac{d^{d}\boldsymbol{q}}{(2\pi)^{d}}|I_{A}(\boldsymbol{q})|^{2}C(\boldsymbol{q}),\quad I_{A}(\boldsymbol{q})\equiv\int_{A}d^{d}\boldsymbol{x}e^{i\boldsymbol{q}\boldsymbol{x}}.
\end{equation}
Consider, for simplicity, the case where $A$ is a ball of radius $\ell$. Then the Fourier transform of the indicator function $I_A$ can be calculated explicitly, reading
\begin{equation}
\label{eq:FormFactorA}
I_{A}(\boldsymbol{q})=\left(\frac{2\pi\ell}{q}\right)^{d/2}J_{d/2}(q\ell).
\end{equation}
It is a rapidly oscillating function for momenta $q \sim 1/\ell$; as we are interested in the asymptotic behavior at $\ell \to \infty$, we can average over these oscillations, yielding:
\begin{equation}
|I_{A}(\boldsymbol{q})|^{2}\approx\left(2\pi\right)^{d}\frac{\ell^{d-1}}{\pi q^{d+1}} 
\end{equation}
and thus
\begin{equation}
C_{\ell}^{(2)}\approx\frac{\sigma_{d-1}\ell^{d-1}}{\pi}\int_{\sim1/\ell}^{\sim1/l_{0}}dq\frac{C(q)}{q^{2}},
\label{eq:C2lCq}
\end{equation}
with $\sigma_{d-1} = 2 \pi^{d/2} / \Gamma(d/2)$ a surface of $d-1$-dimensional unit sphere. Here, we have also introduced an infrared cutoff at the scale $\sim 1 / \ell$, where oscillations are not developed, and the ultraviolet cutoff at the scale $\sim 1 / l_0$, where the NLSM starts to be applicable. For the Gaussian approximation \eqref{eq:CqGaussian-appendix}, we obtain
\begin{equation}
C_{\ell}^{(2)}\approx\frac{g_{0}}{\pi} \sigma_{d-1} \ell^{d-1}\ln\frac{\ell}{l_{0}},
\label{eq:C-A-2}
\end{equation}
which is Eq. (12) 
of the main text. This result can be generalized to a region $A$ of an arbitrary shape with characteristic size $\ell$; the result is given by 
Eq.~\eqref{eq:C-A-2} with  $\sigma_{d-1} \ell^{d-1}$ replaced by the area $S_A$ of the region $A$. 

Going beyond the Gaussian approximation, we note that Eqs. \eqref{eq:CumulantA}, \eqref{eq:FormFactorA} can be rewritten in RG-like form as:
\[
\frac{dC_{\ell}^{(2)}}{d\ln\ell}=(d-1)C_{\ell}^{(2)}+\sigma_{d-1}\ell^{d}\int_{0}^{\infty}dq\frac{C(q)}{q}\frac{d}{d(q\ell)}\left(q\ell J_{d/2}^{2}(q\ell)\right)
\]
The integral converges at $q \ell \lesssim 1$ and can be estimated as $\sim -\ell^d C(r\sim \ell)$. This estimation, together with scaling form $C(r) \sim - G(r) / r^{2d}$, allows us to write an RG-like equation for the cumulant:
\begin{equation}
\frac{dC_{\ell}^{(2)}}{d\ln\ell}=(d-1)C_{\ell}^{(2)}+\alpha G(\ell),
\label{eq:RG-cumulant}
\end{equation}
where $\alpha = O(1)$ is a geometry-dependent constant and $G(\ell)$ the dimensionless conductance governed by RG equation (13) 
from the main text. We speculate that a similar equation can also be written for the entanglement entropy. Higher cumulants in the Klich-Levitov expansion are expected to produce series in inverse powers of $G$ on the right-hand side of the corresponding RG equation, which will combine into some $\beta$-function for the entanglement entropy.

\section{Details of computational approach}\label{sec:supp:computational}

\subsection{Numerical simulation algorithm}

The state of the system was characterized by the $N \times N$ (with $N = L^d$ being total number of sites in the system) correlation matrix $\widetilde{\mathcal{G}}_{\alpha\alpha^\prime}$ in the eigenbasis of the Hamiltonian $\hat{H}$ consisting of states $\psi_\alpha(\boldsymbol{x}) \equiv V_{\alpha \boldsymbol{x}}$ corresponding to eigenenergies $E_\alpha$. The Green function in the coordinate basis is then given by:
\begin{equation}
\label{eq:supp:GF}
    {\cal G}_{\boldsymbol{x}\boldsymbol{x}^{\prime}}=V_{\boldsymbol{x}\alpha}^{\ast}\widetilde{{\cal G}}_{\alpha\alpha^{\prime}}V_{\alpha^{\prime}\boldsymbol{x}^{\prime}} \quad \Leftrightarrow \quad \hat{{\cal G}}=\hat{V}^{\dagger}\hat{\widetilde{{\cal G}}}\hat{V}
\end{equation}

Fixing the total time $T$ of evolution to be sufficiently large to reach the steady state, the total number of measurements $M$ was then drawn from the Poissonian distribution characterized by the expectation value $\left\langle M \right\rangle = \gamma N$, the times of measurements $t_m$ were drawn from the uniform distribution $t_m \in [0, T]$, and their positions $\boldsymbol{x}_m$ were drawn from the discrete uniform distribution of all sites $\boldsymbol{x} \in [1, L]^d$. 

Between consecutive measurements at times $t_m$ and $t_{m+1}$, the matrix $\widetilde{\mathcal{G}}_{\alpha \alpha^\prime}$ underwent unitary evolution according to:
\begin{equation}
\widetilde{\mathcal{G}}_{\alpha\alpha^{\prime}}(t_{m+1})=\widetilde{\mathcal{G}}_{\alpha\alpha^{\prime}}(t_{m})e^{-i(E_{\alpha^{\prime}}-E_{\alpha})(t_{m+1}-t_{m})},
\end{equation}
so that no time Trotterization was required for our setup. For measurement at a given site $\boldsymbol{x}$, the probability of a ``click'' outcome was calculated as $n(\boldsymbol{x}) = v^\dagger \hat{\widetilde{\mathcal{G}}} v$ with $v_\alpha \equiv V_{\alpha \boldsymbol{x}}$. Depending on the random outcome of the measurement, the matrix $\widetilde{\mathcal{G}}$ was updated accordingly. Specifically, for the ``click'' outcome, it was updated as:
\begin{equation}
\hat{\widetilde{\mathcal{G}}}^{\prime}=\hat{\widetilde{\mathcal{G}}}+vv^{\dagger}-\left(\hat{\widetilde{\mathcal{G}}}v\right)\left(\hat{\widetilde{\mathcal{G}}}v\right)^{\dagger} / n(\boldsymbol{x}),
\end{equation}
while for the ``no-click'' outcome, it was updated as:
\begin{equation}
\hat{\widetilde{{\cal G}}}^{\prime}=\hat{\widetilde{{\cal G}}}-vv^{\dagger}+(v-\hat{\widetilde{\mathcal{G}}}v)(v-\hat{\widetilde{\mathcal{G}}}v)^{\dagger}/(1-n(\boldsymbol{x})).
\end{equation}
Our algorithm was then able to perform a single measurement efficiently using just $O(N^2)$ operations.

We have run simulations using the following initial conditions: (i) ground state of the Hamiltonian $\hat{H}$; (ii) random bitstring product state in the coordinate basis; (iii) random bitstring product state in the eigenbasis, and we have carefully checked that the properties of the steady state are insensitive to the chosen initial condition. For each system size and measurement rate, the simulation was repeated $\sim 100$ times to perform averaging over different quantum trajectories. For each quantum trajectory, we have obtained the Green function \eqref{eq:supp:GF}, which was used to calculate the density correlation function utilizing Eq. (26) from the main text. The latter was then averaged over different quantum trajectories and positions in the system, finally yielding the correlation function of interest, $C(\boldsymbol{x}-\boldsymbol{x}^\prime)$, whose Fourier transform is shown in Fig. 1 of the main text. The particle number covariance, which is shown in Fig. 2 of the main text, was obtained using Eq. (23) of the main text.

\subsection{Collapse analysis}
The finite size scaling analysis of the particle-number covariance data shown in Fig. 2 of the main text was performed using the \texttt{pyfssa} package \cite{PyFSSA}, which is based on method proposed in Ref. \cite{Houdayer2004} (see appendix therein). It optimizes the ``quality'' function, which is the reduced $\chi^2$ statistic:
\begin{equation}
\chi^2(\nu,\gamma_c) = \frac{1}{N}\sum_{L,\gamma}\frac{(G_{L,\gamma}-Y_{L,\gamma})^{2}}{dG_{L,\gamma}^{2}+dY_{L,\gamma}^{2}}.
\end{equation}
Here, $G_{L,\gamma}$ and $dG_{L,\gamma}$ denote values and standard errors for each data point (obtained from statistical averaging over different runs of simulation) and $Y_{L,\gamma}$ and $dY_{L,\gamma}$ denote the estimated values and standard error of a ``master curve'' at points $x_{L,\gamma} = L^{1/\nu} (\gamma - \gamma_c)$, which is obtained by considering local weighted least squares approximation of available data \cite{PyFSSAdocs1}. The ``quality'' is then minimized with respect to $\gamma_c$ and $\nu$, using the iterative simplex algorithm proposed by Nelder and Mead \cite{NelderMead1965} to obtain the estimation of the critical exponent and the transition point. The algorithm also allows for estimation of the ``width'' of the minima, as described by the Hessian of the quality function around its minima. The Hessian is then directly related to the covariance matrix of the fitting parameters, whereas the square root of diagonal elements of the covariance matrix gives the estimation of error bars provided in the inset of Fig. 2 and in the part ``Numerical analysis'' of the main text.

\section{Relation between the entanglement entropy and particle-number fluctuations}

In the Gaussian approximation, particle-number cumulants $C_A^{(N)}$ of order higher than second ($N > 2$) vanish. This implies that higher cumulants contain smallness in parameter $g^{-1} \sim \gamma / J \ll 1$ and justifies keeping only the second cumulant in the Klich-Levitov relation [Eq. (6)
of the main text], $\mathcal{S}_E \approx (\pi^2 / 3) C_A^{(2)}$. In $d=1+\epsilon$ dimensions, the measurement transition happens at $\gamma / J \ll 1$, 
which guarantees that the relation $\mathcal{S}_E \approx (\pi^2 / 3) C_A^{(2)}$ holds around the transition, and thus the entropy $\mathcal{S}_E$ exhibits the same scaling behavior as the second cumulant $C_A^{(2)}$. At the same time, in the 2D problem, the transition takes place at $\gamma / J$ of order unity. By virtue of continuity, it is natural to expect that the equivalence of the scaling of $\mathcal{S}_E$ and $C_A^{(2)}$ holds also in 2D. (The opposite would mean that there is some critical dimension $d_c$ satisfying $1 < d_c <2$ where the critical behavior changes qualitatively.)

To verify these analytical arguments numerically, we have performed simulations of a system of size $L \times L$ with $L = 32$ for three different values of $\gamma = 1.5,~2.9,~4.0$ as representatives of ``metallic'', critical and ``insulating'' phases, respectively. For each realization of a quantum trajectory, we have calculated the single-body Green function $\mathcal{G}_{A,\boldsymbol{r}\boldsymbol{r}^{\prime}}=\left\langle \hat{\psi}_{\boldsymbol{r}}^{\dagger}\hat{\psi}_{\boldsymbol{r}^{\prime}}\right\rangle $ (see Sec. \ref{sec:supp:computational}) of a strip $A$ of size $L \times \ell$ (so that $\boldsymbol{r}, \boldsymbol{r}^\prime \in A$) for all possible values of $\ell \in [1, L/2]$. Both the bipartite entanglement entropy and fluctuations of the number of particles were then evaluated by using the relations
\begin{equation}
    {\cal S}_{E}^{(A)}=-\Tr[\hat{{\cal G}}_{A}\ln\hat{{\cal G}}_A+(1-\hat{{\cal G}}_A)\ln(1-\hat{{\cal G}}_A)],\quad C_{A}^{(2)}=\Tr[\hat{{\cal G}}_{A}(1-\hat{{\cal G}}_{A})].
\end{equation}

\begin{figure}[ht]
\centering
\includegraphics[width=0.8\textwidth]{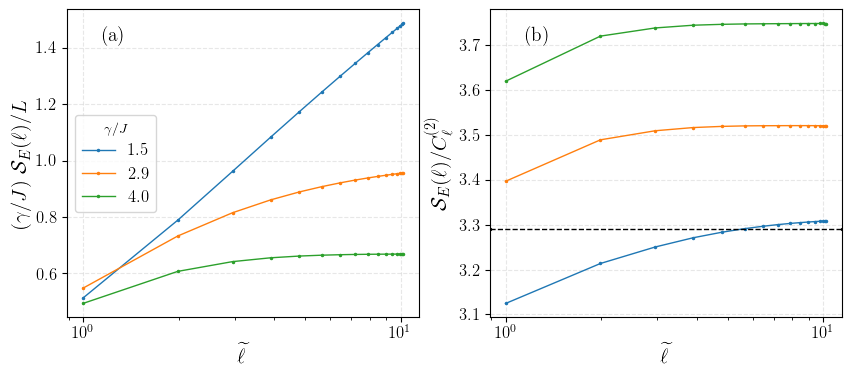}
\caption{Entanglement entropy for a strip of size $L \times \ell$ [here $\widetilde{\ell} = (L / \pi) \sin (\pi \ell / L)$ to take into account finite size effects] for three different values of $\gamma$, corresponding to ``metallic'' ($\gamma = 1.5$), critical ($\gamma = 2.9$) and ``insulating'' ($\gamma = 4.0$) phases. (a) Entanglement entropy (rescaled by $\gamma / J L$). For $\gamma/J = 1.5$, the logarithmic growth is observed (``metallic'' behavior), while for $\gamma/J = 4.0$, one observes saturation corresponding to the ``localized'' (area-law) phase. At criticality, $\gamma/J = 2.9$, there is also saturation, but the approach to it is slow (power-law), at variance with an exponentially fast approach in the ``localized'' phase. (b) The ratio $\mathcal{S}_{E}^{(A)} / C_{A}^{(2)}$ of the entanglement entropy and the second cumulant of number of particles. Saturation is observed for all values of $\gamma/J$ (in the ``metallic'' phase, at criticality, and in the ``localized'' phase), with a saturation value being a smooth function of $\gamma/J$, which remains close to the value $\pi^2/3$ (shown by dashed line) analytically derived for the limit $\gamma / J \to 0$.}
\label{fig:EntanglementEntropy}
\end{figure}

The results are presented in Fig.~\ref{fig:EntanglementEntropy}. The left panel of the figure demonstrates the area-law scaling of the entropy $\mathcal{S}_{E}^{(A)}$  at $\gamma > \gamma_c$ and the ``area law times logarithm'' scaling at $\gamma < \gamma_c$, i.e., exactly the same behavior as that of the cumulant $C_{A}^{(2)}$.
For a full quantitative comparison of both observables, we show in the right panel the $\ell$ dependence of the ratio $\mathcal{S}_{E}^{(A)} / C_{A}^{(2)}$. The key observation is that this ratio saturates, in the large-$\ell$ limit, at a constant value in both phases (``metallic'' and ``insulating'') and at the critical point. Furthermore, the saturation value only weakly depends on $\gamma/J$. At $\gamma/J = 1.5$, it is very close to the value $\pi^2/3$ analytically predicted for $\gamma/J \to 0$. Then it slowly increases when $\gamma/J$ increases and crosses the critical point, reaching the value $\approx 3.75$ at $\gamma/J = 4.0$ (which is well on the ``localized'' side of the transition). 
  
Thus, all the results of this work for the scaling behavior of the second particle-number cumulant $C_{A}^{(2)}$ at the measurement-induced transition in a 2D system equally apply to the entanglement entropy $\mathcal{S}_{E}^{(A)}$.

\section{Scaling analysis in momentum representation}

In the main text, we performed the scaling analysis of the entanglement transition by addressing the correlation function in the real-space representation. In this section of the Supplemental Material, we present more details on manifestations of the transition for $d=2$ in the momentum representation. We recall that $C(q)$ is the correlation function of two conserved currents, and thus should vanish at $q=0$. Then, Eq. (20) from the main text predicts the following behavior of the correlation function for $1 < d < 2$:
\begin{equation}
    C(q)\propto G_{c}q^{d},\quad \text{at}~G_0 = G_c,\quad \text{for}~1 \leq d < 2.
\end{equation}
At the same time, for $d=2$, which is the case of main interest in the present paper, the critical behavior $C(r) \sim -G_c/r^4$, which follows from the scale invariance of the Noether current operator at criticality, predicts a logarithmic correction to this expression:
\begin{equation}
C^{\text{(2D)}}(q)\approx\int C(r)\left(1-\frac{1}{2}q^{2}r^{2}\right)d^{2}\boldsymbol{r}=\frac{1}{2}G_{c}q^{2}\int_{\sim a}^{\sim q^{-1}}\frac{d^{2}\boldsymbol{r}}{r^{2}}\sim G_{c}q^{2}\ln\frac{1}{q a},\quad G_0 = G_c,
\label{eq:Cq-q2logq}
\end{equation}
where the lower cutoff $a$ is the maximum of lattice spacing (unity) and mean free path $l_0$. In the ``localized'' phase, the correlation function decays faster, so that integral $\int C(r) r^2$ converges, leading to $C(q) \propto q^2$ behavior. These two asymptotics should match at $q \sim \ell_{\text{corr}}^{-1}$ in the vicinity of the transition, which predicts:
\begin{equation}
\label{eq:Cq-q2}
C^{\text{(2D)}}(q)\sim G_{c}q^{2}\ln\frac{\ell_{\text{corr}}}{a},\quad G_0 < G_c \,.
\end{equation}
On the other side of the transition, in the ``metallic'' phase, the Gaussian result $C(q) \propto q$ should hold, which yields:
\begin{equation}
\label{eq:Cq-q}
C^{\text{(2D)}}(q)\sim G_{c}q/\ell_{\text{corr}},\quad G_0>G_{c} \,.
\end{equation}

The behavior of $C(q)$ at $q\to 0$ limit thus allows one to distinguish between the different phases and, in particular, to locate the transition. The corresponding numerical data are shown in Fig.~1
of the main text. One clearly sees in that figure that there is a transition between the ``metallic'' behavior, $\lim _{q\to 0}C(q)/q > 0$, and the ``insulating'' behavior, $\lim _{q\to 0}C(q)/q = 0$, with increasing $\gamma$. 

The data for $C(q)$ as shown in Fig.~1 
of the main text are very useful for visualizing the transition. At the same time, it turns out that $C(q)$ is not the optimal observable for an accurate numerical determination of the transition point and of the critical behavior. The reason for this is that the behavior of $C(q)$ at the critical point differs from that in the ``localized'' phase only by a logarithmic factor, see Eqs.~\eqref{eq:Cq-q2logq} and \eqref{eq:Cq-q2logq}. This makes it harder to accurately extract the critical point from the numerical analysis with limited system sizes. In addition, the logarithmic factor complicates the scaling analysis near criticality. This should be contrasted with the particle-number covariance, which has three distinct types of behavior (constant at criticality, linear increase in the ``metallic'' phase, and exponential decay in the ``insulating'' phase), see Eq.~(24)
and Fig. 2
of the main text.

\begin{figure}[ht]
    \centering
    \includegraphics[width=\textwidth]{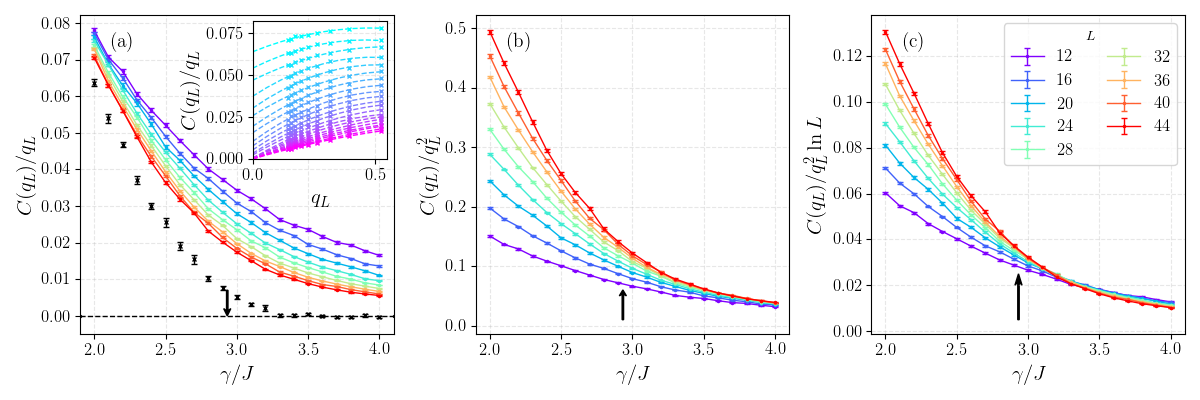}
\caption{Long-wavelength behavior $C(q \to 0)$ evaluated numerically at the lowest available momenta $q_L = 2 \pi / L$ for system sizes from $L=12$ to $L=44$ with periodic boundary conditions. Arrow marks the transition $\gamma_c$ as obtained from the scaling of the particle-number covariance in the main text. 
(a) The ratio $C(q_L) / q_L$ that should be finite in the ``metallic'' phase and zero in the``insulating'' phase at ${q_L\to 0}$. Black points are extrapolated (${q_L\to 0}$) values using quadratic polynomial extrapolation, see inset. 
(b) The ratio $C(q_L) / q_L^2$, which should be $L$-independent in the ``localized'' phase and acquire a logarithmic dependence in the critical point.
(c) The ratio $C(q_L) / q_L^2 \ln L$, which should be $L$-independent at the critical point at $L \to \infty$.}
\label{Sfig:Cq}
\end{figure}

Even though the momentum-space $C(q)$ is not optimal for an accurate determination of $\gamma_c$ and$\nu$,  a closer look at the numerical behavior of $C(q)$ around the critical point is very instructive. The corresponding data are shown in Fig.~\ref{Sfig:Cq}.
Specifically, in this figure, we focus on the lowest possible momenta for available system sizes and present the results for $C(q_L)$ with $q_L = 2\pi / L$ for system sizes from $L=12$ to $L=44$.

Figure \ref{Sfig:Cq}(a) depicts the ratio $C(q_L) / q_L$ (cf.~Fig.~1 
of the main text), which has the meaning of the effective ``conductivity'' in the limit $L \to \infty$ and, according to Eq.~\eqref{eq:Cq-q}, should be non-zero in the ``metallic'' phase and vanish in the ``insulating'' phase. 
We use a quadratic polynomial extrapolation $L \to \infty$. As in Fig.~1 
of the main text, we observe that extrapolated values become indistinguishable from zero at $\gamma \gtrsim 3.2$. While this extrapolation should work perfectly in the metallic and localized phases (i.e., when the system size is larger than the correlation length), it is not accurate at criticality (i.e., in close vicinity of $\gamma_c$), where the behavior of $C(q)$ contains a logarithmic-in-$q$ factor down to the smallest available $q$, see Eq.~\ref{eq:Cq-q2logq}. This leads to a larger finite-size error if one uses this approach for determining the transition point $\gamma_c$. The value $\gamma_c = 2.9$ as obtained by a more accurate approach (see main text) is shown by an arrow. 

In Fig.~\ref{Sfig:Cq}(b), the same data is shown in a different way. Specifically, we plot $C(q_L) / q_L^2$, which is useful to probe the ``insulating'' phase. Indeed, according to Eq. \eqref{eq:Cq-q2}, the ratio $C(q_L) / q_L^2$ should be $L$-independent for system sizes larger than correlation length $L \gtrsim \ell_\text{corr}$ at $\gamma > \gamma_c$. The results are fully consistent with this prediction: one can observe that upon increasing system size $L$, curves on the insulating side converge to a limiting curve. Deviation from this limiting curve becomes visible at a value of $\gamma$ that decreases with increasing $L$,  converging to $\gamma_c$. 

Finally, Fig.~\ref{Sfig:Cq}(c) shows the same data in one more way: here the ratio $C(q_L) / q_L^2 \ln L$ is presented. According to Eqs.~\eqref{eq:Cq-q2logq}, \eqref{eq:Cq-q2}, and \eqref{eq:Cq-q}, it should produce a crossing point at the transition, with an $L$-independent value at $\gamma=\gamma_c$. Indeed, a crossing point around $\gamma \approx 3.2$ is observed. A closer inspection reveals that, upon increasing system size, it drifts towards smaller values of $\gamma$, in full consistency with the actual value $\gamma_c = 2.9$ obtained in the main text. 

Summarizing, the numerically obtained scaling of the correlation function $C(q)$ is fully consistent with the analytical predictions. At the same time, it leads to a larger finite-size error if directly used for extracting the position of $\gamma_c$ for the reasons explained above.

\bibliography{refs}

\end{document}